\documentclass[12pt]{article}
\usepackage{epsfig}
\renewcommand{\title}{        A Speculative Remark on Holography    
}
\advance\textheight by \topskip
\renewcommand{\baselinestretch}{1.2}
\renewcommand{\thefootnote}{\fnsymbol{footnote}}

\newcommand{\beq}{\begin{equation}}
\newcommand{\eeq}{\end{equation}}
\newcommand{\bea}{\begin{eqnarray}}
\newcommand{\eea}{\end{eqnarray}}

\begin{document}
\newtheorem{fig}[figure]{Figure}


\renewcommand{\baselinestretch}{1}
\renewcommand{\thefootnote}{\alph{footnote}}

\thispagestyle{empty}

\vspace*{-0.3cm} {\bf \hfill LBNL--42058}

\vspace*{-0.3cm} {\bf \hfill November 98} \vspace*{1.5cm}
{\Large\bf \begin{center} \title \end{center}}
{\begin{center}

\vspace*{0.5cm}
   {\begin{center} {\large\sc
                Richard Dawid\footnote{\makebox[1.cm]{Email:}
                                  dawid@thsrv.lbl.gov}}

    \end{center} } 
\vspace*{0cm} {\it \begin{center}
    Theoretical Physics Group \\
    Ernest Orlando Lawrence Berkeley National Laboratory \\
    University of California, Berkeley, California 94720, USA
    \end{center} }
\vspace*{1cm}

\end{center}

{\Large \bf \begin{center} Abstract \end{center} }

Holography suggests a considerable reduction of degrees of freedom in theories
with gravity. However it seems to be difficult to understand how holography
could be realized in a closed re--contracting universe. 
In this letter we claim that
a scenario which achieves that goal will eliminate all spatial degrees of
freedom. This would require a different concept of quantum mechanics
and would imply an intriguing increase of power for the natural laws.

\renewcommand{\baselinestretch}{1.2}

\newpage
\renewcommand{\thefootnote}{\arabic{footnote}}
\setcounter{footnote}{0}


\section{Introduction}
\label{intro}

Recently the concept of holography attracted much attention.
Motivated primarily by the area 
law of black hole entropy the idea is that any theory involving gravity
can be fully described by a theory without gravity on the boundary, where the
number of information quanta is one per Planck area on the boundary
\cite{th-s}. The relevance of
this holographic principle is supported by the concept of matrix theory
\cite{bfss}
which suggests a holographic nature and even more so by the discovery
of a duality between string theory on AdS space and a conformal
super Yang Mills theory on its boundary \cite{m-w-p}. 

In this letter we want to deal with two general questions that arise in the 
context of holography. The first question is how holography can be realized in
a closed re--contracting universe. As it was argued in \cite{fs}, a 
straightforward estimate of the entropy in a closed universe is 
irreconcilable with a holographic limit for the number of degrees of freedom.
To make holography consistent with a  closed re--contracting universe, rather
radical changes in the understanding of the situation seem to be necessary.

The second question is an even more fundamental one.
It asks about the meaning of the notion of degree of freedom in principle,
a question that is raised by the advent of holography.

The classical paradigm behind the notion of physical 
degrees of freedom is the following: There exist physical laws which uniquely
define the timely evolution of some
initial state. The space--like situation of this initial state is in principle
unconstrained. The number of degrees of freedom of a physical system
denotes the number of possibilities to construct that initial state. 
In a quantized framework this picture is slightly modified since the 
time--like determinism is reduced by the uncertainty relation. 
However the basic 
conception of degrees of freedom as a consequence of space--like freedom
in contrast to time--like determinism remains the same. 

A holographic principle makes this distinction between determined and
``free'' directions more problematic. It claims that the number of dimensions
spanning the space of degrees of freedom is not directly connected 
to the number of space--like dimensions. In some sense the physical theory
that determines time--like evolution is also able to reduce the freedom of
choice of a space--like initial state. Having started out in this 
direction, one could question the concept of degree of freedom altogether.
If a physical theory is able to abolish some of the spatial degrees of freedom,
why not all of them? 
Physical theories underwent a remarkable evolution so far. 
While classical mechanics is 
merely a tool to describe and predict the kinematics of given objects in a
given space, today's string theory tries to explain 
uniquely all qualitative features
of the universe like types of particles and forces, number of dimensions, etc. 
It seems to be the natural next step to aim at uniquely prescribing as well 
the individual spatial localization of the objects that fill the universe. 
One can argue that a theory of the world only
deserves its name if the local spatial structure of this world is part of 
the theory. We will call a theory that achieves that goal spatially unique.
 
In this letter we claim that the two questions described above are
connected: The holographic principle seems to imply spatial uniqueness 
in the case of a closed re--contracting universe.
In section two we give a simple toy-example of a spatially unique world.
This example serves merely as an illustration and is not an essential part
of the main discussion. The reader who is not interested in it can skip it
without loosing the argument.
Section three discusses the basic arguments that suggest spatial uniqueness
in a realistic cosmological setting. Section four speculates on how quantum
mechanics would have to change in this framework. We end with conclusions.

\section{A very simple example}


At first sight the quest for a spatially unique theory 
might seem to be a rather over--ambitious one.
However, in this chapter we want to argue that even at a purely classical 
level it is not out of reach.
The property we need is naturally present in the currently 
much discussed AdS space, though not too much appreciated there: A closed
time dimension. The discussions of AdS usually use the universal covering
space to avoid this property. In the context of our discussion it will be 
very useful. 

We assume a classical toy universe with a rigid space--time characterized by
a 2--dimensional space--like sphere of radius $R$ and a compactified time 
dimension.\footnote{3 spatial dimensions are just slightly more complicated
due to parity.}
In this space--time we assume a large number $n$ of ideally elastic 
distinguishable balls with average diameter
$d$ and average mass $m$ which move around with a certain total energy and 
bounce into each other. We only count balls which interact at least indirectly 
with the main group of balls. Single balls or interacting subsets of balls 
which do not interact with any ball of the largest interacting subset 
are invisible and therefore ignored.

Each phase space constellation of the $n$ balls is considered one microstate.
We formally treat the number of microstates as a finite number and eventually
send it to infinity. Our model satisfies  a time--like 
boundary condition $F(t=0)=F(t=T)$ with $F(t)$ being the microstate of the 
system at time $t$ and $T$ being the length of the time dimension. 
We want to estimate how many initial 
microstates exist for a given $T$ which are consistent with this condition. 
To be able to do so we have to find scenarios where the number of consistent
initial microstates is still controlled by statistics.
This is not the case if there exists any subset
of microstates whose histories are considerably less complex than the 
genuine system. Statistics of the whole system would be blind against
an increased number of consistent initial states among those microstates.
The basis for the formation of a less complex subset is some symmetry
that splits the system into smaller parts. Three types of less complex 
histories can result: a) States that return to the initial state after a 
fraction of the prescribed time; b) states whose histories always remain in a 
lower dimensional subspace; and c) states that can be split into smaller 
identical sub--ensembles with identical histories.

a) can be excluded by defining the closed time length as the 
time up to the first reappearance of the same microstate. Using that
definition histories that return after a fraction of that time are no 
solution for the chosen time length any more and don't count. 

b) can be excluded by taking
$nd>2\pi R$. In that case the system does not fit into a lower dimensional
subspace.

c) can be excluded by making $n$ a prime number. In that case 
sub--ensembles can only exist if the whole system lives in one dimension 
which has already been excluded above.

There obviously exists a direct connection between 
an initial microstate and its time inverted state.
If one is a consistent initial state, the other is as well, without telling
anything new since we do not have a global time arrow. We will therefore 
count pairs of initial microstates which are the time inverted of
each other and ask how many pairs are consistent initial 
states.\footnote{This counts some initial microstates which include 
non--interacting balls (if they interact in the time inverted case)
and double--counts the time inversion invariant initial microstates. 
However the first don't change statistics and the second are
far too few to be statistically significant.}.

Now we can start counting: 
Our system is characterized by a set G of 2N possible microstates 
$F_i$ at time $t=0$ (a set $G^{\prime}$ of $N$ states plus the $N$ time
inverted states). Each of those microstates develops 
into some state of the same set $G$ at the time $t=T$. Since we considered a 
finite space,  we can take the time period between
$t=0$ and $t=T$ sufficiently long to allow the assumption that each of the 
states in G can be realized with the same probability at the time $T$. 
Now, as explained above, we forget about the time inverted
half of our states. Statistically the probability for one microstate of our 
set $G^{\prime}$ of $N$ states to satisfy $F(t=0)=F(t=T)$ is $N^{-1}$. 
The average number of states $F_i$ of $G^{\prime}$ for which 
$F(t=0)=F_i=F(t=T)$ is one. 
The probability that there exist exactly $n$ microstates that fulfill the 
condition above is 

\bea
P(n)=e^{-1}(1-e^{-1})^n
\eea

The probability that exactly one microstate meets the condition therefore 
is

\bea
P(1)=e^{-1}-e^{-2}
\eea

This means that with considerable probability there are no 
degrees of freedom in such world. The described model is spatially unique 
for a large fraction of choices for the parameters of our model.

An important question is whether our toy model 
can reproduce  the second law of thermodynamics at least locally. 
To discuss this we have to assume that the initial state
has low entropy. Genuinely this is a highly improbable assumption, however
it is necessary for comparing our toy model with more realistic scenarios.
One could formulate the assumption the following way: We have,
depending on the parameters T, N, E/m, R/d and the individual
diameters and masses of the balls, an infinite number of possible
universes. We select the small subgroup of them, whose single
closed history passes through a low entropy state. 
Now we want to discuss how a genuine closed history of a 
universe in this subgroup would look like. We look at the universe 
at a certain time $T_1<<T$. One can assume that the decoupling of the 
boundary condition, i.e. the principle that any state at T can be reached from
$T_1$ with the same probability is still valid. Therefore though the system
is not a truly statistical system, up to $T_1$ it
will genuinely behave like a thermodynamical system, because the same 
statistical arguments that select the correct macrostate in a thermodynamical
system also define the probabilities for the one path that leads back to 
its initial state to sit in some macrostate at $T_1$. It is simply much more 
probable that the ``right'' path is one of the many in the thermodynamically
favoured macrostate than one of the few in some low entropy macrostate.

The argument does not hold any more
if the time period between $T_1$ and $T$ is too small since in that case it
becomes more unlikely for microstates far away from the initial state
to be able to return. The assumption of the same probability for all 
microstates is not valid any more. Since we do not have a 
global time arrow, the whole argument described above holds for an inverted 
time axis as well. Therefore one would naturally 
expect that the universe evolves according to the second law of thermodynamics
during its earlier phase and increasingly deviates from that path 
while approaching the point $t=T/2$.\footnote{This happens if T is too small
to actually reach the maximal entropy. Otherwise the middle phase would be
a maximal entropy period.} The
fundamental laws of thermodynamics are
viable only in the first phase. 
In the middle phase of the universe the ``tendency'' towards the initial
state becomes stronger and entirely destroys the statistical principle. 
The third
phase of the universe shows an inverted entropy law and a negative time arrow
like it was proposed in a cosmological setting in \cite{g}, 
\cite{h}(withdrawn in \cite{hll}) and \cite{kz} in a cosmological setting. 

\section{Cosmology}


So far we have just discussed a simple toy universe to show that spatial
uniqueness is nothing absurd in principle. To make contact with the real
world it is necessary to introduce something like a compactified time 
dimension into a realistic cosmological scenario. 

A closed time can only exist in a 
re--contracting universe since a universe that expands for ever obviously 
does not allow an identification of different points in time. 
$T$ must be identified with the time period between
big bang and big crunch. There exist two pure types of spatially
homogeneous re--contracting universes. The first achieves re--contraction via
a negative cosmological constant. This is the Anti DeSitter universe which
has infinite spatial extension. And the second is a closed universe with an
over--critical mass content. This second case is interesting for our
purposes. From now on ``closed universe'' will always denote 
a closed re--contracting universe.

At this point we come back to the holographic conjecture. 
In spaces other than AdS it is not clear how a geometric picture of
holography should look like
since there is no nice notion of a boundary of space. 
In \cite{fs} it was proposed to identify the boundary with
the horizon of the universe. In that case the question whether or not a
holographic information bound is obeyed for all time is decided by a race 
between the expansion of the universe (that decreases the entropy density)
and the expansion of the horizon (that decreases the ratio between the
horizon area and the volume it encloses). It was estimated that holography 
according to that formulation is in principle consistent with the 
evolution of entropy in an open universe.\footnote{An investigation
that comes to similar conclusions was done in \cite{br} in the framework of
pre-big-bang cosmology.} However it fails in a closed universe
provided one accepts the conventional identification between entropy
and degrees of freedom as well as a ``thermodynamic'' behaviour of the
universe throughout all times. Now these are exactly the principles we were
ready to abolish to gain spatial uniqueness. 

In the following we will not adopt the formulation of holography used in
\cite{fs}. We will just start with a very fundamental requirement that
seems to be necessary for holography to make sense:
If holography is true, the number of possible microstates when the
universe has the size of the Planck length is one. This has to be the case
coming from the big bang as well as going into the big crunch. A holographic
principle that does not enforce that condition would have to use some
``boundary'' beyond the actual spatial realization of the
universe and therefore would loose its original meaning.
If we accept this condition, 
a holographic closed universe somewhat resembles a scenario with a closed
time dimension. One could formally make the identification
by gluing together its Planck--scale sized ends. 

It is interesting to note that the time--symmetric boundary conditions 
enforced by holography show up as well if one tries to formulate a low entropy
boundary condition for the wave function of the universe \cite{hh}.
The wave function of the universe can be formulated as the functional integral

\bea
\Psi(h_{ij})=K\int_C\delta g(x) exp[iS_E(g)]
\label{fi}
\eea

where  $K$ is a normalization factor, $S_E$ is the classical gravity action 
and the integral sums over 
all four geometries that eventually end at a space--like boundary with the 
induced three metric $h_{ij}$. Since time is part of the four--space that
is varied in the variation principle above, different paths need different 
times to reach $h_{ij}$ which means that time cannot be used as a universal 
evolution parameter for $\Psi$. Consequently the state function of $\Psi$
can only be formulated as a time independent Schroedinger equation, the
Wheeler DeWitt equation

\bea
H\Psi(h_{ij})=0
\label{se}
\eea

where $H$ is the gravitational Hamiltonian.
If we want to introduce boundary conditions  for $\Psi$ we cannot use time as
a parameter. We have to use something connected directly to a specific
space--like surface, e.g. the actual extension of the universe. 
This however means
that the boundary conditions can only be imposed simultaneously for a 
certain extension in the expanding as well as the contracting phase. 
The boundary conditions must be time--symmetric \cite{kz}. 

To come back to the consequences of holography, let us sketch how the 
evolution of the universe from minimal initial entropy can take place.
This is nothing unusual and part of the the standard lore of cosmology.
The universe starts from an initial quantum fluctuation which induces
the initial minimal entropy state. An inflationary period blows up the
universe exponentially and after re--heating the universe is in a state
of rather high entropy mainly carried by the photons in thermal
equilibrium. From there on entropy is increased by gravitational clumping
due to small initial density variations. Eventually this gravitational 
clumping will lead to the formation of black holes which carry an entropy per
baryon that is much higher than the photon/baryon ratio. Therefore,
if the universe would behave statistically throughout its evolution, 
black holes would dominate the entropy in the later contracting phase of 
the universe.

Now the next question is, what must happen to return to the minimal
entropy at the final state. We have to note that the 
situation differs fundamentally from the situation in our classical toy model.
There we had the choice to pick exactly the right microstate at $t=0$
and we found that there is approximately one such state. Now we have only 
one possible initial microstate.
Why should this state lead back to itself at $t=T$? 
Let us consider the number of microstates
at $t=T/2$. If our system would be classically deterministic, we could go back
in time and be sure to find always exactly one microstate that is causally
connected to each state at $t=T/2$. The number of microstates would be 
constant. In our universe however this is not the case and the reason lies
in its quantum nature. In
a universe that initially has just one microstate the whole richness of
possibilities emanates from quantum fluctuations. Imagine the universe 
measures itself at $t=T/2$ to be in a specific state. This means that
the chain of factual quantum effects has
led to this state while all the other states
are non--factual hypothetical results of all possible quantum ``decisions'' 
up to that time. Consequently if we want to translate the principle of
the toy model into a realistic scenario, we have to replace the classical
probabilities for specific initial states entirely with quantum probabilities
for specific quantum decisions during the evolution of the universe.
The boundary conditions imposed by holography act as ``teleological''
hidden parameters that determine quantum decisions and guide the way back to
the initial state. 

Now we come to the crucial argument for spatial uniqueness. 
The maximal possible entropy of the universe we can see respectively feel
today, realized if all matter ends up concentrated in a single black hole, 
would be 

\bea
S_{max}=10^{123}~.
\eea

Therefore the finetuning of the
universe to reach minimal entropy in the big crunch is higher than
$exp({-10^{123}})$.
This represents
an immensely specific selection of paths from the set of paths allowed by
conventional quantum mechanics. This extreme selectivity will have a
strong impact on the character of local quantum decisions. 
A specific outcome of a measurement that would be perfectly probable
according to quantum statistics at that time could well imply extremely
improbable future quantum decisions in order to find its way back to the 
holographically determined final state and therefore be virtually excluded. 
The plausibility of a specific quantum decision at a certain time is determined
not only by some wave function at that time but by all future ``probabilities''
the state has to go through to reach the final state.
Thus it is impossible to keep
the interpretation of the wave function as something connected to
a probability density for a specific outcome of a measurement
at a certain time. 

The only way to keep a probability interpretation
of the wave function would be to integrate over the probabilities of a whole
measured path\footnote{It is important to distinguish the measured path
from a trajectory in the path integral. We assume that the universe
continuously measures itself through self interaction and therefore makes a
continuous chain of quantum decisions that define its state at each stage
modulo the fundamental quantum uncertainty. An evolution of the universe
defined by this chain of measurements we call a measured path.} 
throughout the evolution of the universe and understand
the integrated probability as the probability for the physical reality
of a certain measured evolution of the universe. However, 
since there is just one universe, there is no way
to understand the meaning of statistics in that context. 
Statistics only makes sense
if it is applied to a large set of events. It is definitely meaningless to
keep a statistic principle if its validity is reduced to one single event 
at all, namely the overall form of the universe\footnote{This does 
not change at all if one uses the notion of Everett branching. There is no 
philosophically sound way to define a statistical principle that covers
a set of non--interacting parallel universes.}. Using a statistical explanation
for the quantum--dynamics of the universe in this case is indistinguishable 
from not giving any explanation at all.
Therefore the only way to avoid loosing all control over the dynamics of 
the universe is to postulate a  
mechanism that chooses unambiguously one measured path. This however,
following the arguments above, picks uniquely one spatial distribution at
each point in time, in other words it means spatial uniqueness. 

We started off by demanding holography for a closed universe
and we ended up abolishing all spatial degrees of freedom. How does this
fit into a geometrical picture and how can it be compared to the AdS case? 
Holography on
AdS relies on two basic properties:  1) There are no
Cauchy surfaces in AdS. In other words physics on AdS 
does not entirely emanate from an initial state at a time $t=0$ but always
gets new input from the boundary. 2) There is the nonlocal 
character of observables in gravity. This enforces a situation where
the boundary defines physics in the bulk entirely. 

In the case of a closed universe we have again
two basic properties. 1) Cauchy surfaces do exist. 
Thus the whole evolution of the universe emanates from an initial state. 
2) We have seen that gravity, if holographic, uniquely
determines the initial state and its evolution. 
This is to some extent reminiscent
of the AdS case: An initial state in the bulk does not contribute to the 
number of degrees of freedom of the system. But, because of 1) it is more than
that: It implies spatial uniqueness. 
This is also reflected in the geometric picture of a boundary on which 
all information of the enclosed space can be stored. 
If one encloses some space inside a boundary and then maximizes this 
space, in a closed universe the boundary will shrink to a point and 
therefore the information capacity there will be reduced to zero.

We do not see how this type of argument could be adapted to tell anything new 
about holography in eternally expanding universes.

\section{Prospects and problems of a new quantum principle}

We do not have a specific proposal for a deterministic principle that
could lead to a phenomenologically reasonable world. In this section we will 
just discuss some basic properties which should be realized in such a 
concept and mention some important problems.

The essential character of  the required quantum principle must be the 
following: On one side it has to abolish the statistical nature of 
quantum mechanics to satisfy the holographic bound. But on the other side
it has to keep the uncertainty principle to allow an increase of ``possible''
microstates during the evolution of the universe.
This might look rather strange at first sight, however there is no
contradiction in it. While the uncertainty principle is a direct
implication of the mathematical structure of quantum mechanics, rooted in
the non--commutativity of conjugate operators, the identification of
the wave function with the square root of a probability density 
represents an additional interpretative step. While the mathematical
description remains at the level of wave mechanics, the probability
interpretation is based
on the introduction of an additional selection principle, in the Copenhagen
interpretation the concept of contraction of the wave function
in an experiment. Of course the statistical interpretation of the 
wave function is well
founded in our observation of the world, however there is no reason
to believe that this statistical interpretation cannot be replaced by
some new principle that achieves the same phenomenology up to our stage of
the universe. 
Due to their ``teleological'' nature the hidden parameters that select 
the measured path  are surely nonlocal and
do not violate Bell's inequalities. The world is totally deterministic
but not exactly well defined, there remains a core of uncertainty of
observables that is irreducible by experiment.

The most natural approach towards a quantum principle of the described type 
would be to define a pseudo--classical action principle that exactly chooses
the measured path with the highest over all ``probability''. The word
``probability'' in this context is just used for traditional reasons but has
lost its original meaning, therefore it is set under quote. 
Obviously this approach would
be impossible in a world without a final boundary condition.
It would force each quantum decision towards the most probable outcome and
therefore could not provide the observed statistical character of the quantum
world. In a closed universe however the boundary conditions enforce local
decisions of various ``probabilities''. Nevertheless it is not easy to 
imagine how
a realistic phenomenology could be realized. There exist two fundamental
problems: How can an action principle lead to a quantum statistics whose
distribution is controlled exactly by the square of the wave function and
not by the wave function to some arbitrary power. And why does the selected 
path lead to a universe 
that is not ideally isotropic but shows the observed density perturbations.
Even if a scenario could answer those two questions, probably it would still
have to fix a certain maximal extension of the universe as an additional 
boundary condition (analog to the fixing of
a certain duration T in the classical case). Otherwise 
the highest ``probability'' solution should always be the trivial case of 
non--existence. 

What follows are some short remarks on other sensitive points. 
The T--invariant character of the universe in the discussed scenario
requires a T--invariant conception of quantum mechanical measurement that
is not realized in the conventional 
chronological concept of the contraction of the wave function. There exists 
a formulation of quantum mechanics without a time arrow \cite{hh2, gh} 
that can be applied to our framework.

A second remark concerns the return to the same microstate that is 
required in our framework. According to \cite{gh} in a world where 
gravitational effects can be neglected and de--coherence is exactly realized, 
no nontrivial
history leads back to its initial state. In our case however in the crucial
early and late phases of the universe both conditions are not fulfilled
and the argument does not apply.

Another question is whether the effects of a symmetric re--contracting 
universe should have been seen in experiments. In fact there exists an
experimental result \cite{p} that excludes a closed time--symmetric
universe based on the would--be--effects 
of symmetric propagation of star light, 
provided there exist no ``unnatural conspiring relations'' that could hide
this effect. However, as
remarked in several works that consider time--symmetric universes 
(see e.g. \cite{kz}), such unnatural conspiring relations that could e.g. 
enhance photon scattering and tune its direction in a middle phase of 
the universe in any case constitute an essential part of a time--symmetric
universe scenario. Therefore it seems very difficult to make any real
statement based on experiment in that respect.

Finally one has to say that the closed universe scenario has lost some of
its appeal recently because of measurements of distant supernovae which
hint towards an open universe with a positive cosmological constant 
\cite{p-r-g}. Still it is probably too early to draw final theoretical
conclusions from those measurements.

\section{Conclusion}


In this letter we tried to make plausible that a holographic nature of 
the universe seems to imply spatial uniqueness if the universe is closed. 
It remains to be seen whether the required quantum principles
can be formulated in an exact way and can in fact reproduce the 
quantum phenomena and gravitational phenomena of our observed world.
However we want to emphasize once 
more that the consequences of finding a realistic scenario of that type
would be enormously gratifying.
 
A highly attractive picture would emerge.  
The model would eliminate the indeterminism of quantum 
mechanics without local hidden parameters. It would avoid
the problematic notion of increasing entropy in a contracting 
universe. And finally it would establish
spatial uniqueness as a defining condition of the world.
While the qualitative features and the evolution in time with all its 
parameters could be uniquely prescribed by some M--, or other theory,
the fact that the universe is closed together with the
holographic principle would uniquely prescribe the world's concrete 
realization in space. The  notion of reality in the world would be reduced
to the notion of reality of the natural laws.

Seen in this light the obstacles against
reconciling holography with a closed universe in a conventional way
can be understood as a hint towards a role of the
physical laws that is much more powerful than it is estimated today.

\vspace{.5cm}

{\bf Acknowledgments:} We would like to thank Paolo Aschieri for very
useful discussions. This work was supported in part by the Director,
Office of Energy Research, Office of Basic Energy Services, of the 
U.S. Department of Energy under Contract DE-AC03-76SF00098 and in part
by the Erwin Schr\"odinger Stipendium Nr. J1520-PHY.



\parskip=0ex plus 1ex minus 1ex


\end{document}